\documentclass[11pt]{article}
\usepackage{epsfig}
\usepackage{moriond}

\def\Journal#1#2#3#4{{#1} {\bf #2}, #3 (#4)}

\def\NCA{\em Nuovo Cimento}

\def\NIMA{{\em Nucl. Instrum. Methods} A}

\def\PLB{{\em Phys. Lett.}  B}

\def\PRD{{\em Phys. Rev.} D}

\def\APJ{\em Astrophys. J.}
\def\Science{\em Science}
\def\IJMP{{\em Int. Journ. of Mod. Phys.} D}
\def\Nat{\em Nature}
\def\AA{\em Astron. Astrophys.}
\def\RMA{\em Rev. Mod. Astron.}
\def\LRR{\em Living Rev. Rel.}

\def\be{\begin{equation}}
\def\ee{\end{equation}}
\def\bea{\begin{eqnarray}}
\def\eea{\end{eqnarray}}

\begin{document}
\vspace*{4cm}
\title{THE MAGIC EXTRAGALACTIC SKY}

\author{BARBARA DE LOTTO (on behalf of the MAGIC Collaboration) }

\address{University of Udine and INFN, Sezione di Trieste\\
Italy}

\maketitle\abstracts{
The MAGIC telescope, with its 17-m diameter mirror, is currently the largest single-dish Imaging Air Cherenkov Telescope. It is located on the Canary Island of La Palma, at an altitude of 2200~m above sea level, and is operating since 2004. The accessible energy range is in the very high energy (VHE, $E>100$ GeV) $\gamma$-ray domain, and roughly 40\% of the duty cycle is devoted to observation of extragalactic sources. Due to the lowest energy threshold (25~GeV), it can observe the deepest universe, and it is thus well suited for extragalactic observations.
The strategies of extragalactic observations by MAGIC are manifold: long time monitoring of known TeV blazars, detailed study of blazars during flare states, multiwavelength campaigns on most promising targets, and search for new VHE $\gamma$-ray emitters. In this talk, highlights of observations of extragalactic objects will be reviewed.}

\section{Introduction}
One of the major goals of ground-based $\gamma$-ray astrophysics is
the study of VHE
$\gamma$-ray emission from active galactic nuclei (AGN). 
Except for the radio galaxies M87 and Centaurus A (and possibly 3C66B), and the flat-spectrum radio quasar 3C279, all the currently known VHE $\gamma$-ray
emitters in the extragalactic sky are BL Lac objects. 
The sensitivity of the current Imaging Atmospheric Cherenkov Telescopes (IACT)
has recently enabled detailed studies of these sources in the
VHE $\gamma$-rays domain, providing information for 
advances in understanding the origin of the VHE $\gamma$-rays, 
as well as powerful tools for fundamental physics studies \cite{VHEgamma_review}.

The IACT technique \cite{Weekes} uses the atmosphere as a calorimeter to detect
the extensive air shower produced after the interaction of a VHE $\gamma$-ray. 
The charged particles (mainly electrons and positrons) in the air shower produce Cherenkov light
that can be easily detected in the ground with photomultipliers. A Cherenkov telescope uses
a large reflector area to concentrate as much as possible of these photons and focus
them to a camera where an image of the atmospheric cascade is formed. By analysing this image
it is possible to reconstruct the incoming direction and the energy of the $\gamma$-ray. The analysis of
the images is also used to reject the much higher background of cosmic rays initiated showers.

In this paper selected results on extragalactic observations with MAGIC are presented.

\section{The MAGIC telescope}

MAGIC \cite{magic04,cortina05}, located on the Canary Island of La Palma (2200~m
a.s.l.), is currently the largest (17-m diameter) single-dish IACT. 
Due to its large collection area and uniquely designed camera, MAGIC has reached 
the lowest energy threshold (trigger threshold 50--60~GeV at small zenith angles, new trigger for pulsar observations $\sim$ 25~GeV \cite{CrabPulsar})
for $\gamma$-ray emission among the existing terrestrial $\gamma$-ray telescopes.

MAGIC has a sensitivity of $\sim$~1.6\% of the Crab Nebula flux in 50
observing hours. Its energy resolution is about 30\% above 100 GeV and about
25\% from 200 GeV onwards. The angular resolution is $0.1$ deg. 
The MAGIC standard analysis chain is described,
e.g., in Albert et al.\cite{Crab}.  Observations during moderate moonshine enable a
substantially extended duty cycle, which is particularly
important for blazar observations. Parallel optical $R$-band observations are
performed by the Tuorla Blazar Monitoring Program with its KVA 35-cm telescope.

A second MAGIC telescope is being commissioned \cite{Teshima}, which is improving the sensitivity to $\sim$~0.8\% of Crab in 50 hours. 

\section{The propagation and absorption of $\gamma$-rays}

While travelling long distances without deviations in the fields, VHE $\gamma$-rays suffer the absorption losses due to the interaction with the low energy photons from the extragalactic background light (EBL), limiting the distance to the source that could be detected.
The standard process is $\gamma_{VHE} \gamma_{EBL} \to e^+ e^-$ pair production. 
The corresponding cross section \cite{Heitler60} reaches its 
maximum, ${\sigma}_{\gamma \gamma}^{\rm max} \simeq 1.70 \cdot 10^{- 25} \, {\rm cm}^2$, 
assuming head-on collisions, when the background photon energy is $\epsilon(E) \simeq \left(0.5 \, {\rm 
TeV}/E \right) \, {\rm eV}$, $E$ being the energy of the hard (incident) photon. This shows that in the energy interval explored by the IACTs, $50 \, {\rm GeV} < E < 100 \, 
{\rm TeV}$, the resulting opacity is dominated by the interaction with 
infrared/optical/ultraviolet diffuse background photons (EBL), with $0.005 \, {\rm eV} < \epsilon < 10\, {\rm eV}$, corresponding to the wavelength range $0.125 \, {\mu}{\rm m} < \lambda  < 250 \, {\mu}{\rm m}$.

Based on synthetic models of 
the evolving stellar populations in galaxies as well as on deep galaxy counts~(see, for a review, \cite{hauser}), several estimates of the spectral energy distribution (SED) of the EBL have 
been proposed, leading to different values for the transparency of the universe to $50 \, {\rm GeV} 
< E < 100 \, {\rm TeV}$ photons \cite{EBL}; the resulting uncertainties are large. 

Because of the absorption produced by the EBL, 
the observed photon spectrum $\Phi_{\rm obs}(E_0,z)$ is related to the emitted one 
$\Phi_{\rm em}(E(z))$ by 
\begin{equation} \label{eq:flux.tau}
\label{a0}
\Phi_{\rm obs}(E_0,z) = e^{- \tau_{\gamma}(E_0,z)} \ \Phi_{\rm em} \left( E_0 (1+z) \right)~,
\end{equation}
where $E_0$ is the observed energy, $z$ the source redshift and ${\tau}_{\gamma}(E_0,z)$ is the optical depth \cite{Fazio-Stecker}.

\begin{figure}
\centering
\includegraphics[width=.4\textwidth]{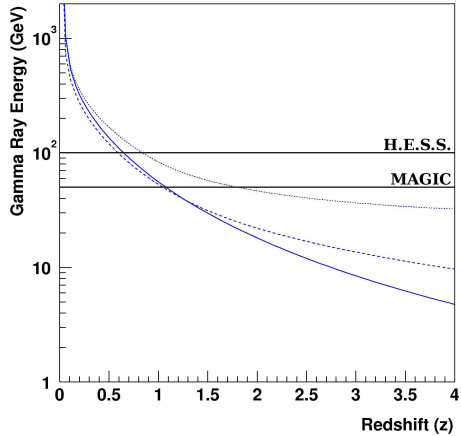}
\caption[]%
{\label{fig:gr-horizon}
Gamma-ray horizon compared with the lower energy limit of
the MAGIC and H.E.S.S.\ Cherenkov telescopes;
the curves of the photon energy versus horizon are computed for different background evolution 
models by Blanch \& Martinez in Ref.~\cite{blanch-martinez}.
}
\end{figure}

The energy dependence of $\tau$ leads to appreciable modifications of the observed source 
spectrum (with respect to the spectrum at emission) even for small differences in~$\tau$, due to the 
exponential dependence described in Eq.~(\ref{eq:flux.tau}).
Since the optical depth (and consequently the absorption coefficient) increases with energy, 
the observed flux results steeper than the emitted one.
The {\em horizon} (e.g.\ Ref.~\cite{blanch-martinez,mannheim1999}) for a photon of energy $E$ is defined as the distance 
corresponding to the redshift~$z$ for which $\tau(E,z)=1$, which gives an attenuation by 
a factor $1/e$ (see Fig.~\ref{fig:gr-horizon}). MAGIC has the lowest energy threshold, and thus is currently the best suited telescope to look farther away.

\section{Multi-Wavelength Campaigns}
Coordinated simultaneous multi-wavelength observations, 
yielding spectral energy distributions (SED) spanning over 15 decades in energy, 
have been recently conducted, and turn out to be essential for a deeper understanding of blazars. 
MAGIC participated in a number of multiwavelength-campaigns on known northern-hemisphere
blazars, which involved the X-ray instruments {\it Suzaku} and
{\it Swift}, the $\gamma$-ray telescopes H.E.S.S., MAGIC and VERITAS, and
other optical and radio telescopes.
\begin{itemize}
\item Mkn\,421 was detected in two campaigns during outbursts in 2006 and 2008; the coordinated effort allowed
 for truly simultaneous data from optical to TeV energies, and studies of correlations between the different energy bands 
 \cite{Mkn421a,Mkn421b}.
\item The VHE emission of PG\,1553+113 showed no variability
during the first multi-wavelength campaign on this blazar in July
2006 \cite{reimer,pg1553a}; it was observed simultaneously for the first time together with AGILE during 2008 \cite{pg1553b}.
\item 1ES\,1959+650 showed VHE data among the lowest flux state
observed from this object, while at the same time a relatively
high optical and X-ray flux (both Swift/Suzaku) was found
\cite{1es1959}. The SED could be modeled assuming a one zone SSC model, using parameters similar to the ones
needed for the SED measured in 2002.
\item Also campaigns on 1ES\,1218+304 and 1H\,1426+428 have been
carried out, during both of which significant X-ray variability
has been observed. The VHE data are being analyzed.
\end{itemize}
Further campaigns have been and will be organized in the future.

\section{Strong Flaring of Messier\,87 in February 2008}

M\,87 is the first non-blazar radio galaxy known to emit VHE
$\gamma$-rays, and one of the best-studied extragalactic black-hole
systems. To enable long-term studies and assess the variability
timescales and the location of the VHE emission in M\,87, the
H.E.S.S., MAGIC and VERITAS collaborations established a regular,
shared monitoring of M\,87 and agreed on mutual alerts in case of
a significant detection. 
During the
MAGIC observations, a strong signal of 8\,$\sigma$ significance was
found on 2008 February $1^{st}$, triggering the other IACTs as well as
{\it Swift} observations. 
The analysis revealed a variable (significance: 5.6\,$\sigma$)
night-to-night $\gamma$-ray flux above 350 GeV, while no
variability was found in the 150--350 GeV range \cite{m87a}. 
The $E>730\,\mathrm{GeV}$
short-time variability of M\,87 reported by \cite{hessm87} has been confirmed. 
This fastest variability $\Delta t$ observed so far in TeV $\gamma$-rays 
in M\,87 is on the order of or even below one day, suggesting the core of M\,87 as the origin of the TeV $\gamma$-rays.
M\,87 is the first radio galaxy that shows evidence for a connection between simultaneously and well sampled radio and VHE flux variations, opening a new avenue for the study of AGN accretion and jet formation \cite{m87b}.

\section{Blazars Detected during Optical Outbursts}

MAGIC has been performing target of opportunity observations upon
high optical states of known or potential VHE $\gamma$-ray emitting extragalactic sources. Up to now,
this strategy has been proven very successful, with the detection of Mkn
180 \cite{mkn180}, 1ES 1011+496 \cite{1es1011}, and recently S5\,0716+71 \cite{s50716} (paper in preparation). 

In April 2008, KVA observed a high optical state of the blazar
S5\,0716+71, triggering MAGIC observation, which resulted in a
detection of a strong 6.8\,$\sigma$ signal, corresponding to a flux
of $F_{>400\mathrm{GeV}} \approx 10^{-11} \mathrm{cm}^{-2}
\mathrm{s}^{-1}$. 
The MAGIC observation time was 2.6\,h. The source was also in a
high X-ray state \cite{atel1495giommi}.

The determination of the before-unknown redshifts of 1ES\,1011+496
($z=0.21$) \cite{1es1011} and S5\,0716+71 ($z=0.31$) \cite{nps08} makes these objects the third-most and second-most
distant TeV blazars after 3C\,279, respectively.

\section{The region of 3C66A/B}
The MAGIC telescope observed the region around the distant blazar 3C~66A for 54.2\,h in August--December 2007. 
The observations resulted in the discovery of a $\gamma$-ray source 
centered at celestial coordinates R.A. = $2^{\mathrm{h}} 23^{\mathrm{m}} 12^{\mathrm{s}}$ and decl.$=43^{\circ} 0.'7$ (MAGIC~J0223+430), coinciding with the nearby radio galaxy 3C~66B \cite{3c66ab}. 
The energy spectrum of MAGIC~J0223+430 follows a power law with a normalization of $\left(1.7\pm 0.3_{\mathrm{stat}} \pm 0.6_{\mathrm{syst}} \right)\times10^{-11}$ 
TeV$^{-1}$ cm$^{-2}$ s$^{-1}$ at 300\,GeV and a photon index $\Gamma = -3.10 
\pm 0.31_{\mathrm{stat}} \pm 0.2_{\mathrm{syst}}$. 
A possible association of the excess with the blazar 3C~66A and nearby radiogalaxy 3C~66B is discussed in these proceedings \cite{mazin3C66}.

\section{Detection of the flat-spectrum radio quasar 3C\,279}
\begin{figure}
\centering
\includegraphics[width=0.6\linewidth]{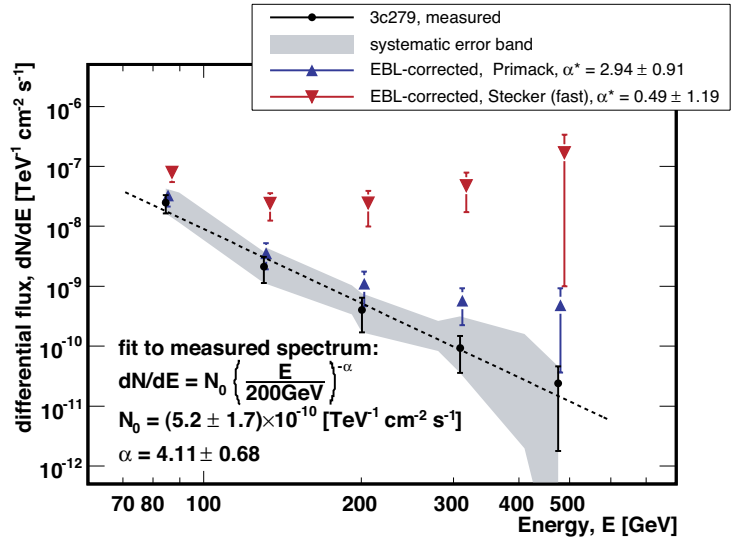}
\caption[]%
{\label{fig:spectrum-3c279} Spectrum of 3C~279 measured by MAGIC.
The grey area includes the combined statistical ($1\sigma$) and
systematic errors, and underlines the marginal significance of
detections at high energy. The dotted line shows compatibility of
the measured spectrum with a power law of photon index
$\alpha=4.1$. The blue and red triangles are measurements
corrected on the basis of the two models for the EBL density.}
\end{figure}

Observations of 3C~279, the brightest EGRET AGN \cite{wehrle},
during the WEBT multi-wavelength campaign \cite{webt} revealed a
5.77\,$\sigma$ post-trial detection on 2006 February $23^{rd}$ supported
by a marginal signal on the preceding night
\cite{MAGICScience}. The overall probability for a
zero-flux lightcurve can be rejected on the 5.04\,$\sigma$ level.
Simultaneous optical $R$-band observations by the Tuorla
Observatory Blazar Monitoring Program revealed that during the
MAGIC observations the $\gamma$-ray source was in a generally high
optical state, a factor of 2 above the long-term baseline flux,
but with no indication of short time-scale variability at visible
wavelengths. The observed VHE spectrum can be described by a power
law with a differential photon spectral index of
$\alpha=4.1\pm0.7_\mathrm{stat}\pm0.2_\mathrm{syst}$ between 75
and 500 GeV (Fig.~\ref{fig:spectrum-3c279}). The measured integrated flux
above 100~GeV on February $23^{rd}$ is $(5.15\pm 0.82_\mathrm{stat}\pm
1.5_\mathrm{syst}) \times 10^{-10}$ photons cm$^{-2}$ s$^{-1}$.

\begin{figure}
\centering
\includegraphics[width=0.6\linewidth]{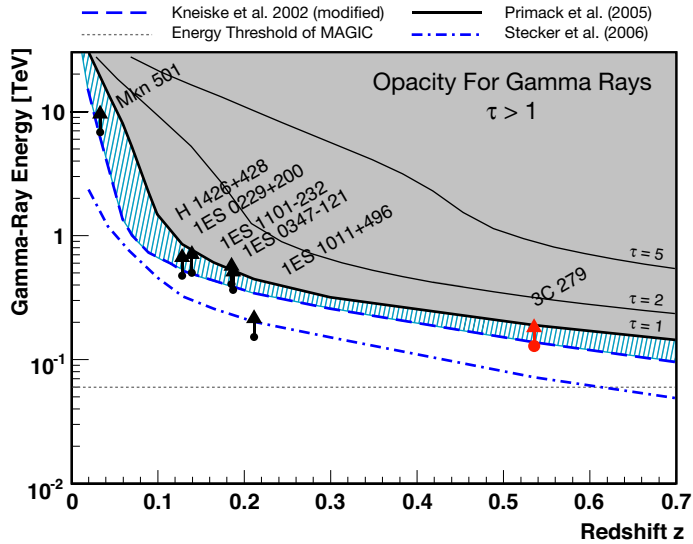}
\caption[]%
{\label{fig:GRH-3c279} The $\gamma$-ray horizon. The redshift
region over which it can be constrained by
observations has been extended by MAGIC up to z=0.536.
}
\end{figure}

This detection extends the test on the transparency of the universe up to $z=0.536$;  the $\gamma$-ray horizon together with the IACT measurements is shown in Fig.~\ref{fig:GRH-3c279} from \cite{MAGICScience}. 

VHE observations of such distant sources
were until recently impossible due to the expected strong
attenuation of $\gamma$ rays by the EBL, which influences the
observed spectrum and flux, resulting in an exponential decrease
with energy and a cutoff in the $\gamma$-ray spectrum. 
The reconstructed intrinsic spectrum is difficult
to reconcile with models predicting high EBL densities, while low-level models,
e.g. \cite{EBL}, are still viable. Assuming a maximum
intrinsic photon index of $\alpha^\ast = 1.5$, an upper EBL limit
is inferred, leaving a small allowed region for the EBL.

\begin{figure}
\centering
\includegraphics[width=.6\textwidth]{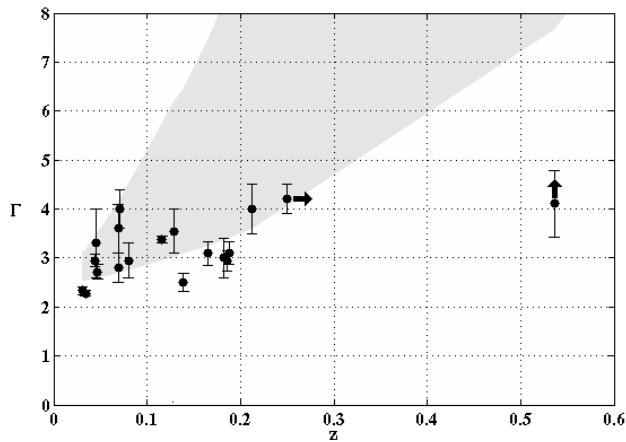}
\caption[]%
{\label{fig:Gamma_zeta-new}
Observed values of the spectral indexes of all the blazars detected so far
in VHE band as a function of the redshift; the grey band represents the prediction for different EBL models.}
\end{figure}

In Fig.~\ref{fig:Gamma_zeta-new} the observed values of the spectral indexes of the blazars detected so far
in VHE band are shown, together with the prediction (light grey area) of the standard scenario. The recent findings suggest a higher transparency of the universe to VHE photons than expected from current models of the EBL, and could be interpreted in terms of more exotic scenarios \cite{DARMA}.

\section{The July-2005 Flares of Mkn 501}
Mkn 501 ($z=0.034$) is known to be a strong and variable VHE
$\gamma$-ray emitter. MAGIC observed Mkn~501 for 24 nights during
six weeks in summer 2005. In two of these (one with
moon present), the recorded flux exceeded four times the
Crab-nebula flux, and revealed rapid flux changes with doubling
times as short as 3 minutes or less. For the first time, short
($\approx$~20 min) VHE $\gamma$-ray flares with a resolved time
structure could be used for detailed studies of particle
acceleration and cooling timescales.
In addition, a time delay between different energy bins could be investigated,
and gave some hints of a delay of the higher energy photons  \cite{alb07a}.

An energy-dependent speed of photons in vacuum is expected as a generic signature in some approaches to Quantum Gravity (QG) theories, where Lorentz invariance violation is a manifestation of the foamy structure of space-time at short distances.
It could be reflected in modifications of 
the propagation of energetic particles, i.e. dispersive effects due to a non-trivial refractive 
index induced by the fluctuations in the space-time foam \cite{Amelino-Camelia}.
The dependence of the speed of light on the energy $E$ of the photon can be parameterized as

\begin{equation} \label{eq:c(E)}
c' = c \left[ 1 \pm \left( \frac{E}{E_{\rm S1}} \right) \pm 
       \left( \frac{E}{E_{\rm S2}} \right)^2  \pm ... \right] \, .
\end{equation}
The energy scales $E_{\rm S1}$, $E_{\rm S2}$ are usually 
expressed in units of the Planck mass, $M_P \equiv 1.22 \times 10^{19}$~GeV/c$^2$. If the linear 
term dominates,  Eq.~(\ref{eq:c(E)}) reduces to
\begin{equation}
c' = c \left[ 1 \pm \left( \frac{E}{E_{\rm S1}} \right) 
\right] \, .
\end{equation}
A favored way to search for such a dispersion relation is to compare the arrival times of photons of different energies arriving on Earth from pulses of distant astrophysical sources (see \cite{Mattingly} for a review).

 The reanalysis of the Mkn~501 data in \cite{alb07b} resulted in a
much-improved estimate of the time-energy relation. At a
zero-delay probability of $P=0.026$, a marginal time delay of
$\tau_l = (0.030\pm0.012)\,\mathrm{s\,GeV}^{-1}$ towards higher
energies was found using two independent analyses, both exploiting
the full statistical power of the dataset (see \cite{me08}
for details).

Since it is not possible to exclude that this delay is due to some energy-dependent
effect at the source, because the emission mechanisms are not currently understood, a lower
limit of $E_{S1} > 0.21 \times 10^{18}$~GeV (95\% ~c.l.) can be
established. 
However, if the emission mechanism at the source were understood and the observed delays were mainly due to propagation, this number could turn into a real measurement of $E_{\rm S1}$.

This pioneering study demonstrates clearly the potential scientific value of an analysis of multiple flares from different sources.

\section{Conclusions}

After almost $4$ observation cycles, MAGIC observations of the extragalactic TeV $\gamma$-ray sources contributed to many physics insights, confirming the rich potential of VHE $\gamma$-ray astrophysics.
Among the currently detected $27$ VHE $\gamma$-ray emitters, MAGIC has discovered 8 new sources, and detected and studied 5 known ones. 

\begin{figure}
\centering
\includegraphics[width=.8\textwidth]{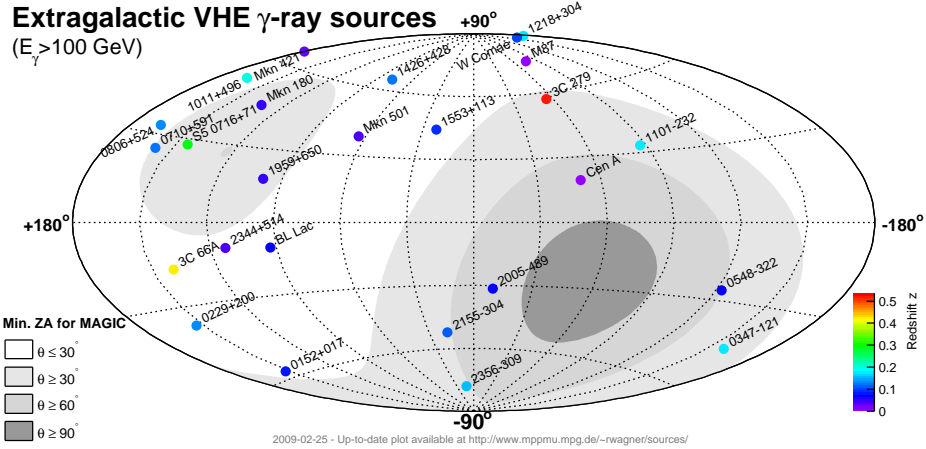}
\caption[]%
{\label{fig:TeV_Skymap}
Skymap of extragalactic VHE $\gamma$-ray sources together with the MAGIC field of view.
}
\end{figure}

In Fig.~\ref{fig:TeV_Skymap} the skymap of the detected sources, together with the MAGIC field of view \cite{skymap}, is shown (see this reference also for an up-to-date list).

Important contributions to the understanding of active galactic nuclei have been given, allowing both to infer the intrinsic properties of the sources and to probe the nature of photon propagation through cosmic distances.

\section*{Acknowledgments}
We thank the Instituto de Astrofisica de Canarias for the excellent working conditions at the 
Observatorio del Roque de los Muchachos in La Palma. The support of the German 
BMBF and MPG, the Italian INFN, the Spanish MCINN, the ETH Research Grant 
TH 34/043, the Polish MNiSzW Grant N N203 390834, and the YIP of the 
Helmholtz Gemeinschaft is gratefully
acknowledged.

\end{document}